\documentclass[11pt]{cernrep}
\usepackage{graphicx}
\usepackage{here}
\begin{document}
 \title{COVERAGE OF CONFIDENCE INTERVALS FOR POISSON STATISTICS IN PRESENCE OF SYSTEMATIC UNCERTAINTIES}
\author{J. Conrad, O. Botner, A. Hallgren, C. P. de los Heros}
\institute{High Energy Physics Division, Uppsala University, Sweden.}
\maketitle
\begin{abstract}
In this note we consider coverage of confidence intervals calculated with and without systematic uncertainties. These calculations follow the prescription originally proposed by Cousins \& Highland but here extended to account for different shapes, size and types of systematic uncertainties. Also two different ordering schemes are considered: the Feldman \& Cousins ordering and its variant where conditioning on the background expectation is applied as proposed by Roe \& Woodroofe. \\
Without uncertainties Feldman \& Cousins method over-covers as expected because of the discreteness of the Poisson distribution. For Roe \& Woodroofe's method we find under-coverage for low signal expectations.\\
When including uncertainties it becomes important to define the ensemble for which the coverage is determined. We consider two different ensembles, where in ensemble A all nuisance parameters are fixed and in ensemble B the nuisance parameters are varied according to their uncertainties. We also discuss the subtleties of the realization of the ensemble with varying systematic uncertainties.
\end{abstract}

\section{INTRODUCTION}
The incorporation of systematic uncertainties in the calculation of confidence intervals is to some extent an unsolved problem. A semi-Bayesian method which has a satisfactory intuitive behavior and is relatively easy to compute has been suggested by R. Cousins \& V. Highland in 1992 [1]. Despite attempts [2], at present no satisfactory solution exists within a completely frequentist framework. 
Coverage is the defining property of confidence intervals calculated with frequentist methods. Whereas frequentist methods should give correct coverage by construction, Bayesian methods which fulfill the requirement of correct coverage are usually much easier to accept by the particle physics community. Therefore, coverage in context of Cousins \& Highland (CH) type methods is of interest.\\
Coverage is only a meaningful quantity if it is defined in terms of an ensemble. A usual definition is (see e.g [3]):\\
\\
{\it An algorithm is said to have the correct {\it coverage} if it
provides intervals which contain the ``true'' value of the 
tested parameter in a fraction 1- $\alpha$ of infinitely many 
{\bf identical} experiments (independent of the tested parameter)
}
\\
\\
\noindent
However, in presence of systematic uncertainties it seems practical to relax this definition by substituting {\it \bf identical} by {\it \bf similar}. Similar here means that the experiments are selected randomly from a distribution of experiments with varying nuisance parameters. \\
In the next section we will describe an extension of the CH method and show how coverage may be calculated for an ensemble of identical experiments. We will then discuss coverage calculations for the case of varying nuisance parameters. The final section will be devoted to discussion and conclusions, in particular we will point to a subtlety of defining the ensemble with varying nuisance parameters. 

\section{THE GENERALIZED COUSINS \& HIGHLAND METHOD}
In the following we will consider the calculation of confidence intervals for the parameter, $s$, of a Poisson distribution with known background, $b$.
The main idea of the CH method is to include systematic uncertainties by integrating over a probability density function (PDF) describing these uncertainties.The required ordering scheme is then applied with a PDF which results from this integration.\\
Consider for example (as done in their original paper) an uncertainty, $\sigma_{\epsilon}$, in the signal efficiency, $\epsilon$. Then the PDF used in the confidence belt construction is modified to:
\begin{eqnarray}
q(n|...) = 
\frac{1}{\sqrt{2\pi \sigma_{\epsilon}^2}}\intop_0^{\infty}  d\epsilon'\, P(n|(b+ s\epsilon'))\,e^{\frac{- (\epsilon - \epsilon')^2}{2\sigma_{\epsilon}^2}}
\end{eqnarray}
The Gaussian describes the a posteriori probability for the ``true'' efficiency $\epsilon'$, given the measured (or assumed) efficiency $\epsilon$.\\
Background uncertainties can be taken into account in a similar way. In the presence of an uncertainty, $\sigma_b$, in the prediction of the level of background processes, the PDF becomes:
\begin{eqnarray}
q(n|s + b) = \frac{1}{\sqrt{2\pi\sigma_{b}^2}}\intop^{\infty}_0
db'\, P(n|s + b')
e^{\frac{-(b-b')^2}{2\sigma_b^2}}
\end{eqnarray}\\
In general both uncertainties described above and an additional uncertainty in the background efficiency are present. Correlations between them might further complicate 
the problem.

\subsection{POissonian Limit Estimator: POLE}
POLE is a FORTRAN 77 program [4] which implements the extended CH method described above. It performs a frequentist construction of confidence belts  and uses a Monte Carlo method to perform the integrals. At present following types of uncertainties are included:\\
\\
{-   Uncertainty in the signal and background efficiencies and correlations.} \\{-  Uncertainty in the level of background processes.}\\
\\
The following possible forms of PDFs describing the uncertainties are included:\\
\\
{-   Gaussian (truncated) }\\
{-   Flat (symmetric, asymmetric)}\\ 
{-   Log - normal (mean/peak centered on 1)}\\
\\
The user can choose between the following ordering schemes:\\
\\
{-   Neyman [5]}\\
{-   Neyman with Conditioning }\\ 
{-   Feldman \& Cousins (FC) [6]}\\
{-   FC with Conditioning (the Roe-Woodroofe (RW) method) [7]} \\
\\
Table 1 shows the FC and RW intervals for two examples of observed events and expected background and for different values of the uncertainty in signal efficiency. The intuitive and generally observed effect of  including systematic uncertainties is a widening of the confidence intervals.  Here one example was chosen explicitly to illustrate the counterintuitive behavior of the Feldman \& Cousins intervals when there are significantly less events observed than background predicted. The interval does not become wider with increasing uncertainties, but narrower. Interestingly, this is not the case for the RW intervals. \\

\begin{center}
Table 1: Examples for intervals with Gaussian systematic uncertainties. $n_0$ denotes the number of observed events, $b$ the expected background.
\vskip0.2cm
\begin{tabular}{|l|l|l|l|l|}\hline
$n_0$ 	&$b$ &signal efficiency uncertainty  & FC        & RW \\ \hline
6     	&12      &0           &{\bf 0; 1.65}  & 0; 4.00\\
     	&        &0.2         &{\bf 0; 1.50}  & 0; 4.20\\
      	&        &0.4         &{\bf 0; 1.35} & 0; 5.15\\
        &        &            &                   &        \\
11      &12      &0           & 0; 5.80              & 0; 6.90 \\     
        &        &0.2         & 0; 5.85              & 0; 7.30 \\
        &        &0.4         & 0; 6.65              &0; 8.95\\\hline
\end{tabular}
\end{center}

\section{COVERAGE CALCULATION FOR STRICTLY IDENTICAL EXPERIMENTS}
In case of considering an ensemble of strictly identical experiments, i.e. with the nuisance parameters fixed, the distribution of experimental outcomes (number of observed events) will follow a Poisson distribution, similar to the case without uncertainties. The coverage can then be calculated using the Poisson distribution (see e.g. [8] for a recent example of this method):
\begin{eqnarray}
(1 -\alpha(s))_{calc} = \sum_{n|s \, \epsilon[s_1,s_2]}P(n|s+b)
\end{eqnarray}
The sum runs over all $n$ which give an interval which contains the true value of $s$. 
For the coverage calculations a variety of assumptions on magnitude, shape of the distribution and type of uncertainties have been considered. The results are exemplified in figures 1 and 2. Figure 1 shows the coverage obtained without uncertainties and with 40 \% uncertainty\footnote{meaning $\sigma$ in case of a Gaussian distribution and Log-normal distribution and half width in case of the flat distribution} in the signal efficiency  for different assumptions about the shape of the PDF. Figure 2 shows the same for Roe \& Woodroofe intervals.
\begin{figure}
\begin{center}
\includegraphics[width=10cm]{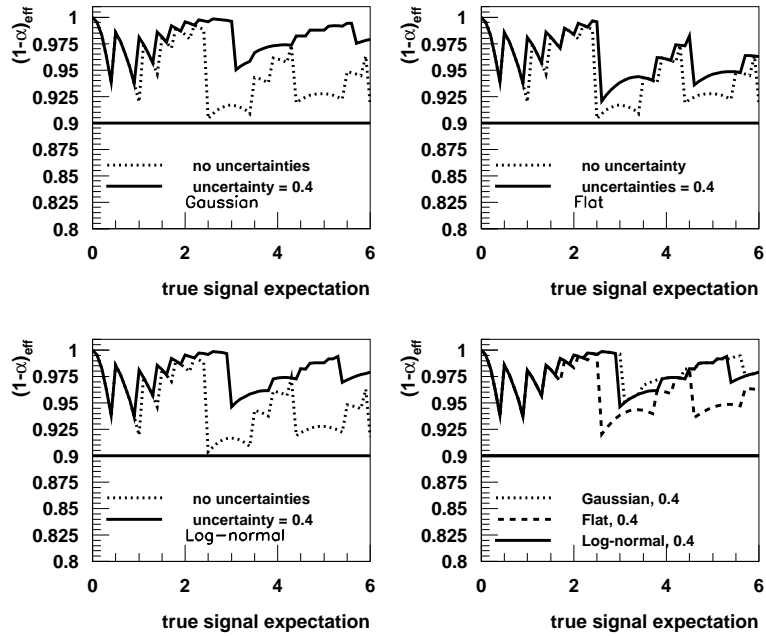}
\caption{Coverage of FC intervals for various assumptions on signal efficiency uncertainties. The coverage obtained with uncertainties is compared to the one obtained with zero uncertainties. The plot on the lower right compares the coverage obtained for different shapes of PDF describing the uncertainties, assuming their magnitude to be 40 \%.}
\end{center}
\end{figure}
\begin{figure}
\begin{center}
\includegraphics[width=10cm]{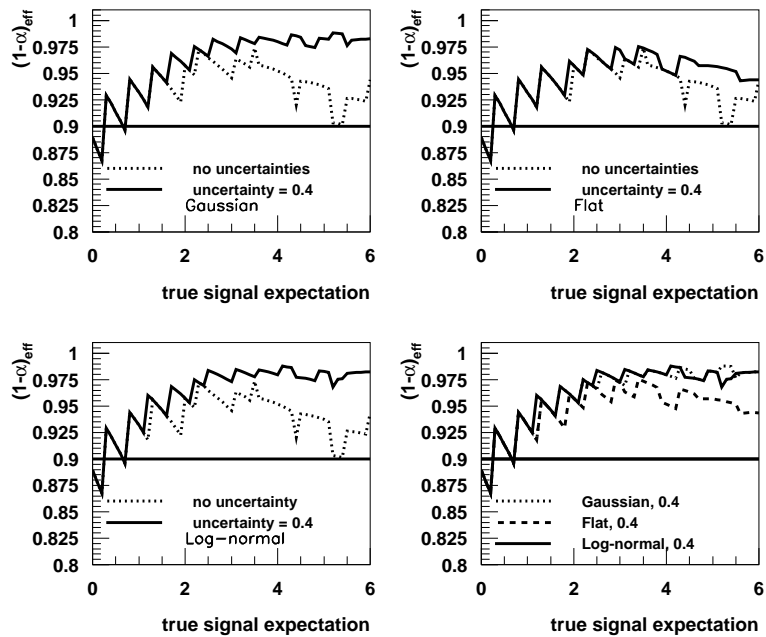}
\caption{Coverage of RW intervals for various assumptions on signal efficiency uncertainties. The coverage obtained with uncertainties is compared to the one obtained with zero uncertainties.}
\end{center}
\end{figure}
The coverage of the intervals without uncertainties is interesting in itself. As also mentioned in [9] the RW intervals undercover for certain signal expectations even in the case without uncertainties, whereas FC is over-covering for all considered values of the signal expectations.
With uncertainties included, the results are similar independent of the shape of the uncertainty distributions. Generally, there is an additional over-coverage as consequence of introducing the systematic uncertainties. The effect is strongest for the Gaussian and log-normal distributions whereas it is somewhat weaker for uniformly distributed uncertainties. \\

\section{COVERAGE CALCULATION WITH MONTE CARLO EXPERIMENTS}
In this section we will consider a definition of an ensemble of repeated experiments more suitable in the presence of systematic uncertainties, viz. similar experiments. The ensemble is realized by Monte Carlo experiments, where for each experiment:
\begin{itemize}
\item{ the efficiency, $\epsilon'$, is randomly selected from a Gaussian distribution (see equation 1)}
\item{ for each trial $\mu_s$, $\epsilon'\, \mu_s$ is taken as mean of a Poisson distribution, $P(n_s, \mu_s)$, from which the ``measurement'', $n_0$, is selected.}
\end{itemize}
As opposed to the case described in section 3, the PDF used in the coverage test is consistent with the PDF used in the calculation of confidence intervals. Thus, per construction, we do not expect to introduce additional coverage. This is confirmed in figure 3.  
\begin{figure}
\begin{center}
\includegraphics[width=10cm]{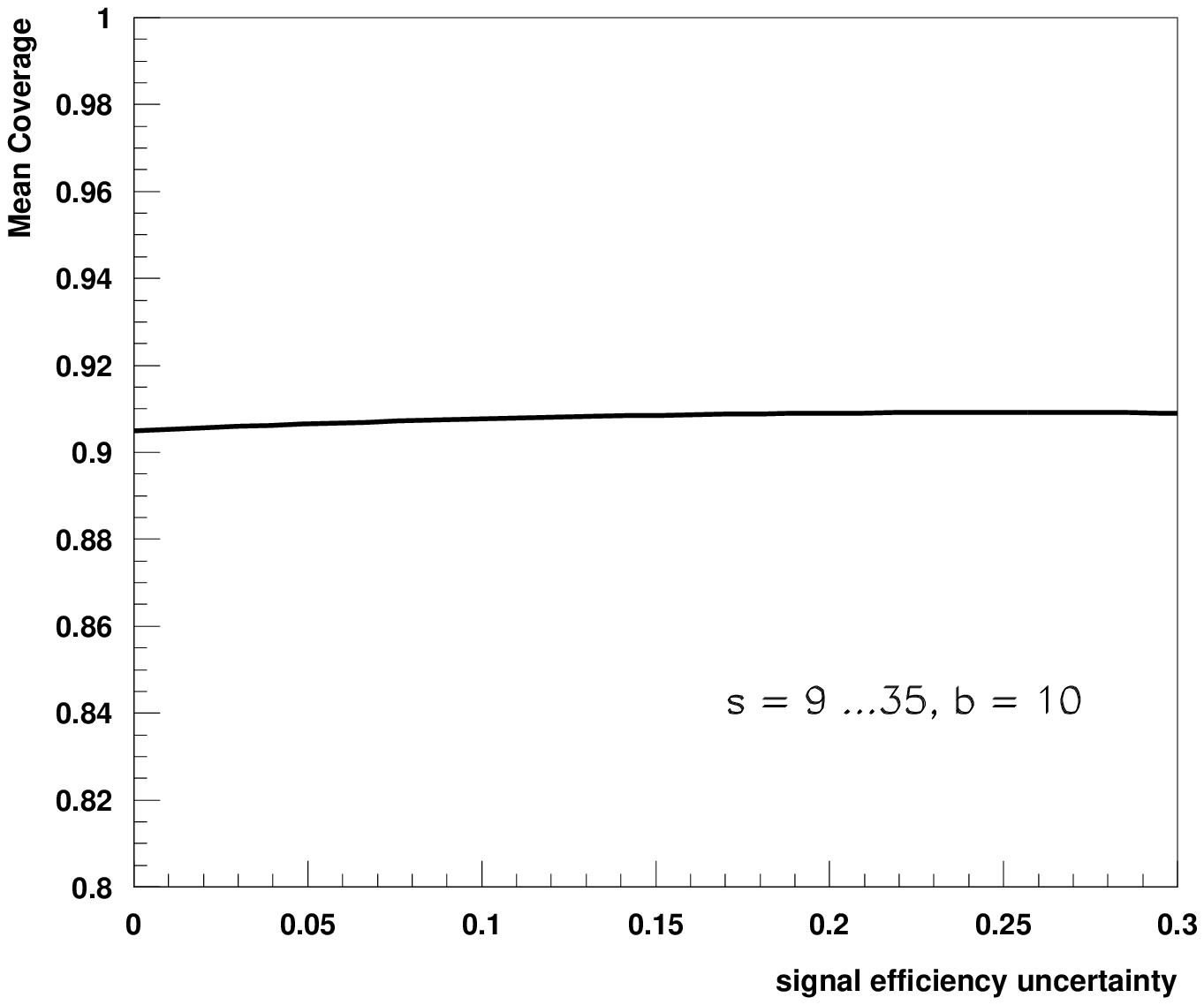}
\caption{Mean Coverage calculated from an ensemble with varying signal efficiency. Mean here taken for a signal expectation range between 9 and 35 and constant background expectation of $b = 10$.}
\end{center}
\end{figure}

\section{DISCUSSION \& CONCLUSION}
As a side remark, it can be argued that the ensemble considered in section 3 (strictly identical experiments) can be interpreted as an ensemble which is conditioned with respect to an ancillary parameter, in this case the efficiency. The applicability of the conditionality principle [9], in this context is beyond the scope of this contribution. However, it has been established in this note, that the conditional coverage of the CH method is not the same as the coverage for the unconditioned ensemble considered.\\
It appears to be generally accepted that, in presence of systematic uncertainties, one should consider an ensemble with varying nuisance parameters. There are , however, different ways of realizing this ensemble in the coverage test. We present here one possibility of varying the nuisance parameters: picking a new ``true'' efficiency every single experiment. In real life, this in principle implies rebuilding the detector for each of the repetitions. With respect to this ensemble, the CH method is in agreement with frequentist statistics. There is no additional over-coverage introduced. \\
However, a typical ensemble of experiments encountered in particle physics is not one where the {\it true} efficiency ($\epsilon'$ in equation 1) varies from experiment to experiment, but instead the measurement of the efficiency ($\epsilon$ in equation 1). To realize this ensemble is computationally more cumbersome, since the construction of confidence belts has to be repeated in each experiment. There are indications that with respect to such an ensemble the HC method over-covers [11]. Preliminary tests we made for two signal/background expectations (true signal equals 3 and 6, no background) support that. If including 40 \% uncertainties in signal efficiency, the coverage increases from $\sim$ 92 \% to $\sim$ 94 \% for required 90 \% confidence level. This seems to indicate a rather modest additional overcoverage. However, conclusive results require a more detailed study of coverage for a range of signal/background expectations and systematic uncertainties. 
\vskip1cm
\noindent

\section*{ACKNOWLEDGEMENTS}
We would like to thank the conference organizers  
Louis Lyons, James Stirling and Mike Whalley for organizing this beautiful
conference. Thanks to Robert Cousins for stimulating discussions.

\end{document}